**Agreeing to disagree, some ironies, disappointing scientific practice and a call for better: reply to "The poor performance of TMM on microRNA-Seq"**

Mark D. Robinson[1,2,*]

[1]Institute of Molecular Life Sciences, University of Zurich, Winterthurerstrasse 190, CH-8057 Zurich, Switzerland

[2]SIB Swiss Institute of Bioinformatics, University of Zurich, Zurich, Switzerland

[*] to whom correspondence should be addressed (mark.robinson@imls.uzh.ch)

**Background.** This letter is a response to a Divergent Views article entitled "The poor performance of TMM on microRNA-Seq" (Garmire and Subramaniam 2013), which was a response to our Divergent Views article entitled "miRNA-seq normalization comparisons need improvement" (Zhou et al. 2013). Using reproducible code examples, we showed that they incorrectly used our normalization method and highlighted additional concerns with their study. Here, I wish to debunk several untrue or misleading statements made by the authors (hereafter referred to as GS) in their response. Unlike GS's, my claims are supported by R code, citations and email correspondences. I finish by making a call for better practice.

**We will agree to disagree on some aspects, while some claims are just untrue**. Scientists often disagree and this is important for the furthering of science. I will reiterate a couple of important arguments from my perspective, since my interpretation of their original results contrasts to theirs. They explain that their original paper focused on "evaluation with different metrics but not downstream analysis". To me, these two go hand-in-hand. Is it even useful to have a metric for a "method" that cannot be used downstream? In particular, the issue of zero counts should not be disregarded and it is not clear how they intend for quantile and lowess normalization to be used in the presence of zero counts. Their claim that "MA plots, and miRNA-qPCR comparisons cannot be done with zero counts" is false; special treatment is given to zero counts for MA plots, as described in the edgeR and DESeq R/Bioconductor packages (Anders and Huber 2010; Robinson et al. 2010), among others, and statistical models for count data readily accommodate zero observations. Why can these comparisons not be done? In addition, many of the metrics are performed on M-values (log-ratios). It is well known that the log transformation *introduces* higher variance at the low expression strength for count data (Robinson and Oshlack 2010; Law et al. 2013). The statistics of differential expression take this into account, but raw M-value comparisons do not, so in fact M-value metrics may inadvertently overemphasize low expression values. Regarding our ROC analyses, GS mention that we "seem to have much smaller numbers" and suggested this may "obscure the AUC interpretation". This may be true; our analyses are based on the 73 miRNAs that have both count-based M-values and qPCR ΔΔCts and the R code is available. I welcome an explanation from GS that explains why their analyses is to be trusted and ours is not, since they have not responded to our requests for more information and have not made code available; at the very least, our analyses highlight how sensitive their analyses to the choics made. Lastly, GS claim that we have "subsequently revised" our TMM method. This is also not true; the *method*, in terms of calculating normalization factors over and above the depth of sequencing, has not changed. We have introduced a convenience function in the software to give easier access to depth-and-normalization-factor-adjusted counts and we have changed the internal representation of the normalization factors. However, we showed in our Divergent Views Supplement that their normalization factors are off by *exactly* the amount of the relative scaling factor, so this change did not affect their original calculations. The change in implementation *does* affect their latest calculations, but I already explained this modification to them in our email exchanges (**Supplementary Note 1**). So, why do GS publish results that they know are incorrect?

**Two striking ironies and further untruths**. There are two gleaming ironic comments made in the GS response. First, regarding performance comparisons with samples from *different* experimental conditions instead of biological replicates, they state "we assume that the majority of the data in experimental conditions are not drastically different when there are no biological replicates, which is reasonable from the MA plots." Yet, in their original manuscript, they *rely* on the fact that there are significant differences to make ROC comparisons, in combination with qPCR validation data. So, the differences in expression are <u>small</u> enough to be ignored comparing performance on M-values, but large enough to used for ROC assessments. I find this a shaky argument. Second, in their response, GS point to code from an edgeR user manual. The *overwhelming* irony here is that, had they followed this example, they would end up with correct TMM-adjusted M-values (See additional R code at **Supplementary Note 2**). However, they did not follow this code. They also did not follow the advice I gave in my original email correspondence (See **Supplementary Note 3**). Admittedly, I did not directly show how the adjustment is done; I hoped it was evident from the code. GS state that "We therefore used the normalization factors as described by Robinson and colleagues", but this is incorrect. What "then-available information" have GS followed? Amazingly, they have not followed the now-available information (e.g. **Supplementary Note 1**) either.

**My real disappointment and an alternative interpretation.** Maybe I am just naïve, but science to me is about the search for truth and ideally a truth that can be reproduced, supported by defensible arguments and that stands the test of time. Unfortunately, GS seem more interested in resolving themselves of any software misuse than presenting a honest representation of the available methods' performance. The senior author commented to me in an email: "It is clearly my intent as is Dr. Garmire's to correct any errors if they were caused". Despite this sentiment, GS do not appear to be interested that their results have errors and are different to mine. Specifically, Figures 1 and 2 of their Divergent Views were sent to me in September 2012 and I responded shortly after (including the *RNA* editor) to explain that their code did not follow our example online and that they misinterpreted our original instructions: dividing the count data and dividing the library sizes by normalization factors are *opposite operations* (**Supplementary Note 1**). Of course, I fully recognize my own irony here. Despite the efforts to correct the miRNA-seq normalization, TMM is only marginally better than total-count normalization (See **Figure 1**). However, given that some methods have never been used for sequencing data and do not have a clear path to downstream analysis, an alternative conclusion is that TMM is perhaps among the best methods available.

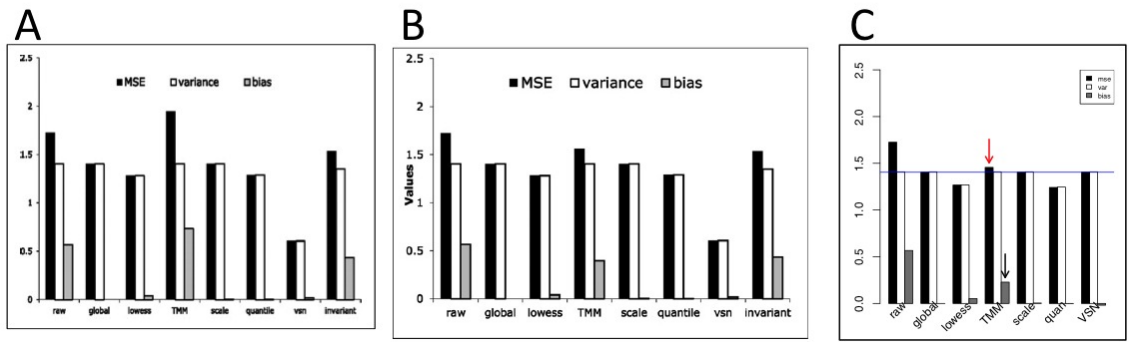

**Figure 1. A) Original incorrect GS analysis. B) Updated incorrect GS analysis. C) Our corrected re-analysis.**

**A couple of recommendations.** I have a couple of recommendations for making these interactions smoother in the future. First, vague questions do not (often) attract good answers, so I encourage people to ask questions in a clear and engaging way. For example, provide a snippet of code, give a plot that shows an unusual result, give ancillary information such as software version and overall, make it easy for the responder to know what you are trying to achieve (without giving sensitive information away). The reason I did not respond *directly* to their original query, is that the business of where to put the normalization factors gets admittedly confusing. I had hoped that pointing to a reproducible example would be a good strategy, but in hindsight, this exposes us to misinterpretation. I am not discouraged by this experience though; I will continue to make code available. Second, software and recommendations may change over time and it is important to follow the instructions that go with the version of software being used. This requires special care, since the web is awash with old user guides and code that no longer runs correctly. Third and most importantly, data analysts should make code available for all analyses and comparisons, potentially including code to recreate every figure (and table) of every paper. This is by no means a new concept (Morin et al. 2012; Ince et al. 2012) and journals are beginning to take additional editorial measures (Editorial 2013). The tools available to do this, especially for R, are readily available (e.g. Sweave, github). That way, data analysis is not a black box.

**Acknowledgements**. MDR acknowledges funding from the European Commission through the 7[th] Framework Collaborative Project RADIANT (Grant Agreement Number: 305626) and SNSF project grant (143883).